\documentclass[%
reprint,
superscriptaddress,
showkeys,
showpacs,preprintnumbers,
 amsmath,amssymb,
pra,
floatfix,
longbibliography
]{revtex4-1}

\usepackage{graphicx}%
\usepackage{dcolumn}%
\usepackage{bm}
\usepackage{color}
\date{\today}
\RequirePackage{lineno}

\begin{document}

\title{Minimizing coherent thermal conductance by controlling the
periodicity of two-dimensional phononic crystals}

\author{Yaolan Tian}
\author{Tuomas A. Puurtinen}
\author{Zhuoran Geng}
\author{Ilari J. Maasilta} \email{maasilta@jyu.fi }
 \affiliation{Nanoscience Center, Department of Physics, University of Jyvaskyla, P. O. Box 35, FIN-40014 Jyvaskyla, Finland}


\begin{abstract}
Periodic hole array phononic crystals (PnC) can strongly modify the phonon dispersion relations, and have been shown to influence thermal conductance coherently, especially at low temperatures where scattering is suppressed. One very important parameter influencing this effect is the period of the structure. Here, we measured the sub-Kelvin thermal conductance of nanofabricated PnCs with identical hole filling factors, but three different periodicities, 4, 8, and 16 $\mu$m, using superconducting tunnel junction thermometry. We found that all the measured samples can suppress thermal conductance by an order of magnitude, and have a lower thermal conductance than the previously measured smaller period, 1 $\mu$m and 2.4 $\mu$m structures. The 8 $\mu$m period PnC gives the lowest thermal conductance of all the samples above, and has the lowest specific conductance/unit heater length observed to date in PnCs. In contrast, coherent transport theory predicts that the longest period should have the lowest thermal conductance. Comparison to incoherent simulations suggests that diffusive boundary scattering is likely the mechanism behind the partial breakdown of the coherent theory.   
\end{abstract}



\maketitle

\section{Introduction}

Engineering of phonon thermal transport is a topic with a wide current interest, with applications such as improving the heat dissipation out of electronic devices \cite{Volz}, reducing the parasitic phonon thermal conductance channel in thermoelectric devices \cite{minnich}, using heat in phononic information processing \cite{Li} and increasing the sensitivity of low-temperature bolometric detectors \cite{Maasilta2016,Rostem}. Traditionally, reduction of thermal conductivity is realized by the introduction of disorder, impurities, nanoparticles, rough surfaces etc. \cite{cahill}, increasing the diffusive scattering of thermal phonons. At sub-Kelvin temperatures, for example, simply roughening the surfaces of a thin plate was predicted to lead to ultralow thermal conductance relevant for ultrasensitive radiation detector applications \cite{puurtinenAIP2014,Maasilta2016}, or at room temperature, thin native oxides at the surface can increase the scattering significantly \cite{neogi}. On the other hand, an interesting question is whether thermal conductance can also be modified by changes in the dispersion relations, or the band structure of phonons \cite{Zen,maldovan}.  

One way to achieve a strong modification of the phonon dispersion is by artificially introducing periodic modification of the elastic properties of the materials in the length scale corresponding to the phonon wavelength range of interest\cite{sigalas,kush}. These structures are known as phononic crystals (PnC) \cite{PnCbook,PnCbook2}, and depending on the contrast between elastic parameters and dimensionality, even complete bandgaps can form at certain frequency ranges, due due to Bragg interference \cite{PnCbook,PnCbook2}. One clear choice for a PnC structure in two dimensions is thus a periodic array of holes, where the simple geometry of a square array of circular holes already creates strong effects \cite{pennec}. Such geometries have been applied in studies of thermal conductivity mostly at room temperatures and somewhat below \cite{tang,yu,hopkins,alaie,anuf1,wagner,lee,Mairee1700027} or above \cite{graczy}, and with periodicities in the 10 nm to 1 $\mu$m length scale. The full theoretical understanding and interpretation of such experiments is still debated and ongoing, mostly because of the sensitivity to sample details \cite{PhysRevB.89.205432}, the varying mean free paths of different wavelength phonons in the broad thermal spectrum \cite{PhysRevLett.107.095901,PhysRevLett.110.025901,regner,PhysRevB.87.214305} and the complex interplay between incoherent and coherent scattering \cite{PhysRevB.87.195301,PhysRevB.89.205432}. It is certainly clear that coherent effects are not guaranteed at high temperatures and at 100 nm - 1 $\mu$m length scales, as the wavelengths of dominant phonons at room temperature are typically $\sim 10$ nm. Recent experiments comparing ordered and disordered hole arrays indeed seem to indicate that coherent effects wash out above 10 K at PnCs of period 100 nm - 300 nm \cite{lee,Mairee1700027}.    

To solve some of these complications, we have shown that at sub-Kelvin temperatures, coherent effects can be dominant in hole array PnCs \cite{Zen}, and that thermal conductance can be reduced by over an order of magnitude with a 2.4 $\mu$m period PnC with a hole filling factor 0.7. Further studies with a 4 $\mu$m period structure  showed even stronger reduction \cite{Maasilta2016}. In addition, coherent effects were also predicted to strongly modify sub-Kelvin heat capacity \cite{TuomasC}. Coherent effects are possible because, first of all, bulk phonon scattering is suppressed \cite{wolfe,klitsner}, and thermal phonons can travel ballistically for very long distances in thin (1 $\mu$m and less) membranes with smooth surfaces, up to distances $\sim 0.1 - 1$ mm \cite{holmes,hoevers,jenni}. Secondly, the dominant thermal wavelengths are increased by three orders of magnitude compared to room temperature, and can reach values above 1 $\mu$m. Moreover, the high frequency phonons are completely frozen; for example, 99 \% of phonons in silicon nitride have frequencies below 20 GHz, and such low frequency phonons have been experimentally shown to retain coherence in hypersonic acoustic wave experiments \cite{wagner} and in Brillouin scattering measurements for thermally populated phonons \cite{PhysRevB.91.075414}. Thus, sub-Kelvin temperature phonons are ideally suited for coherent modification of thermal conductance, and it was already demonstrated that micron-scale periodic structures can have a strong effect \cite{Zen}.  

It has also become clear that a possible phononic bandgap is not the main reason of reduction of thermal conductance in the 2D hole array structures, but it is the modifications of the density of states and group velocity that are dominant \cite{Zen,cryst6060072} and that the period strongly modifies both \cite{Zen,cryst6060072,PhysRevB.93.045410}. Specifically, the band structures are modified in such a way that with increasing period the average group velocity decreases significantly, leading to the counterintuitive effect that the thermal conductance decreases with increasing period (increasing neck size), in contrast to incoherent boundary scattering models where conductance increases with neck size \cite{Lim,Verdier}. The coherent theory \cite{cryst6060072} does not predict a saturation or reversal of this effect up to periods of  8 $\mu$m. However, in real samples with a non-zero diffusive boundary scattering, an interplay between coherent and incoherent phonon populations could lead to an eventual reversal of the trend. 
Thus, there should be a practical limit giving the minimum thermal conduction as a function of the periodicity of the PnC.

Here, we measured the sub-Kelvin thermal conductance of two-dimensional square lattice hole array phononic crystal structures made from 300 nm thick silicon nitride membranes, similar to devices in Refs. \cite{Zen,Maasilta2016} but with much longer periods of 4 $\mu$m, 8 $\mu$m and 16 $\mu$m, keeping the hole filling factor constant at 0.7 (Fig. \ref{samplefig1}). We observe that out of those structures, the 8 $\mu$m period has the lowest thermal conductance, showing that the model of fully coherent and ballistic phonon transport starts to break down beyond that periodicity. The numerical modeling performed suggests that diffusive surface scattering at the hole side walls is likely responsible for this breakdown.   In addition, we compare two PnC structures with the same 8 $\mu$m period but a different heater length, and see that a ballistic model for the heat exchange between the heater and the thermometer can explain the observations of how the measured thermal conductance depends on the heater/thermometer geometry.

\begin{figure}[h]
\includegraphics[width=1\columnwidth]{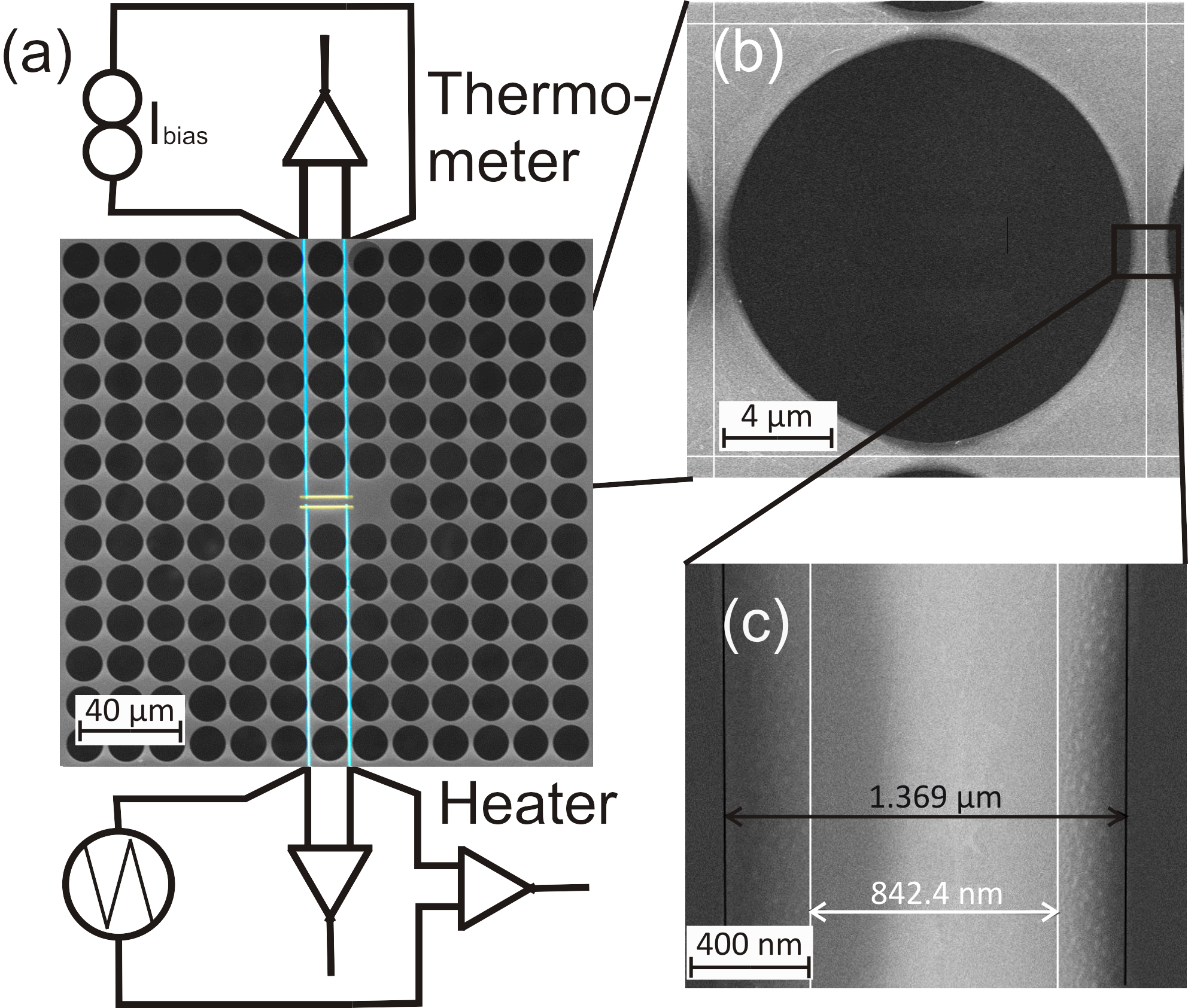}
\caption{Scanning electron micrographs of 2D PnC structures with a periodicity of 16 $\mu$m. (a) Larger scale image of the PnC showing the central region  with the heater and thermometer (yellow), with superconducting leads (blue) extending up and down. The measurement setup is also schematically shown. (b) A Zoom-in of one hole. (c) A zoom-in of a neck region between two holes. The dimensions show the width of the top (white) and bottom (black) neck regions. Their difference indicates sloping sidewalls, with roughness on them visible.}
\label{samplefig1}
\end{figure}

\section{Theoretical Modeling}

Before we show the experimental results, we discuss the theoretical modeling. The band structures of 4 $\mu$m, 8 $\mu$m and 16 $\mu$m period hole array  2D phononic crystals, corresponding to the experiment, were calculated by solving the 3D elasticity equations for an isotropic material \cite{graff} with SiN parameters (Young's modulus $E = 250$ GPa, Poisson ratio 
$\nu = 0.23$, and density $\rho = 3100$ kg/m$^3$) using the finite element method (FEM). Periodic Bloch-wave boundary conditions were used with 2D wave-vectors \textbf{\textit{k}} in the x-y plane, with typically $\sim$ 500 - 1000 k-points in  the irreducible octant of the first Brillouin zone (BZ). Depending on the value of the PnC period, $\sim 16 000 - 87 000$ eigenvalues were computed, extending the frequency range at least to 40 GHz for all periodicities. Unperforated membrane results were also calculated,  with the help of the Rayleigh-Lamb theory \cite{graff,thomas}. 

\begin{figure*}[!htp]
\includegraphics[width=2\columnwidth]{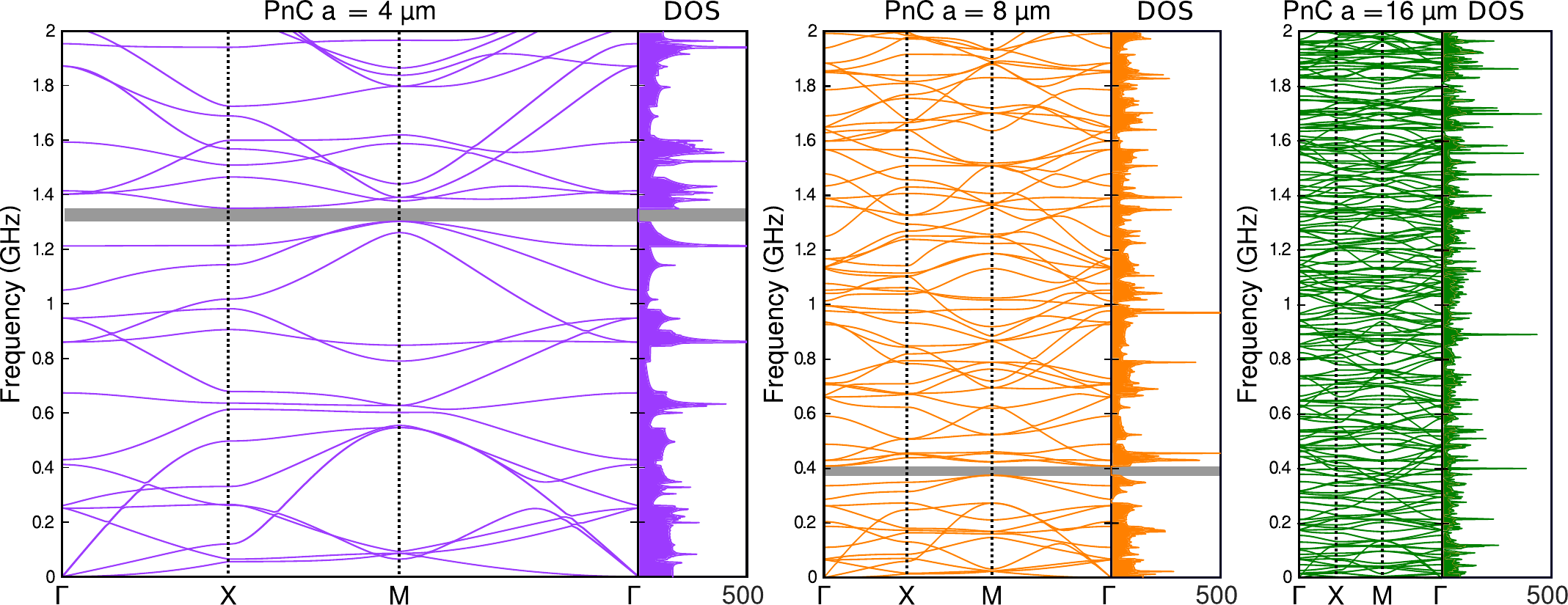}
\caption{\label{spectrumyaolan} Densities of states and dispersion relations  (band structure) in the main symmetry directions of the Brillouin zone (BZ) for the SiN square lattice hole array PnC of thickness 300 nm and hole filling factor 0.7, with periodicity (left) $a= 4 \mu$m,  (middle) $a= 8 \mu$m, and (right) $a= 16 \mu$m.  Complete bandgap is observable at $\nu= 1.3$ GHz  for the PnC with $a= 4 \mu$m, and at$\nu= 0.4$ GHz  for the PnC with $a= 8 \mu$m.} 
\end{figure*}

In Fig. \ref{spectrumyaolan} we compare the  dispersion relations of the three PnC structures, shown in the symmetry directions $\Gamma$-X-M-$\Gamma$ only up to a frequency $\nu =$ 2 GHz for clarity. Full 2D bandstructures were computed for thermal conductance calculations. It is evident that there are great differences between the band structures: the number of bands increases with the periodicity, but the bands become also much flatter. The 4 $\mu$m (8 $\mu$m) period PnC has a complete band gap at 1.3 GHz (0.4 GHz), but its effect is expected to be minor as it only occupies a small frequency region compared to the full thermal bandwidth, which is around 20 GHz even at 0.1 K. The flatness of the bands  will affect both the group velocities $\partial \omega/\partial k$ and the density of states (DOS) (shown in Fig. \ref{spectrumyaolan}), but at this scale of periodicity, the main effect to coherent thermal conduction comes from the reduction of the average group velocity \cite{cryst6060072}.

To calculate the phonon emission from the heater, we take the outward propagating phonon modes (mode index $j$ and wave vector $\textbf{\textit{k}}$) with energies $\hbar\omega_j(\textbf{\textit{k}})$, and derive \cite{cryst6060072,Zen} the following relation for the radiated power  for both PnCs and uncut membranes:

\begin{equation}
\label{PT}
P(T) = \frac{1}{(2\pi)^2}\!\!\sum_j\!\!{\oint_\gamma\!\!{\textrm{d}\gamma}\!\!\!\int_K\!\!\!{\textrm{d}\textbf{\textit{k}} \hbar\omega_j(\textbf{\textit{k}})n\frac{\partial\omega_j}{\partial \textbf{\textit{k}}}\cdot \hat{n}_\gamma \Theta\left(\frac{\partial \omega_j}{\partial \textbf{\textit{k}}}\cdot \hat{n}_\gamma\right)}} 
\end{equation}

where $\gamma$ is the heater element boundary, $\hat{n}_\gamma$ an outer unit normal of that boundary (in the membrane plane) and $\Theta$ is the Heaviside step function. Here $n=n(\omega,T)$ is the Bose-Einstein distribution describing the (assumed) phonon thermal occupation of the emitted phonons and $\partial\omega_j/\partial \textbf{\textit{k}}$ the group velocity of each mode. The only unknown is thus the set of dispersion relations $\omega_j := \omega_j(\textbf{\textit{k}})$ for the permitted phonon modes $j$ in the membrane. The 2D $K$-space integration extends over all $K$-space for uncut membranes, but is replaced by integration over the first Brillouin zone for the PnCs.  Eq. \ref{PT} also assumes that the membrane is infinite so that there is no backscattering from the membrane-bulk interface, and that the bath is at $T_{\textrm{bath}} = 0$ K, so that there is no direct backflow term. The backflow from the bath is taken into account in calculations when comparing with the experiments, by writing $P_{\textrm{net}}=P(T)-P(T_{\textrm{bath}})$, where $P(T_{\textrm{bath}})$ is the calculated power at $T_{\textrm{bath}}$.   

Notice that due to the group velocity term, it is {\em not} possible in general to simplify the above equation to be just an integral over energies weighed by the density of states. This is because the PnC structures are not isotropic so that the group velocity is not a constant on a constant energy surface.

\section{Experimental Results and Discussion}

The measurement was performed with a heater and a thermometer pair at the center of the PnC structure (Fig. 1), by measuring how the temperature of the thermometer rises with the dissipated power at the heater (emitted phonon power). This was achieved by using two normal metal-insulator-superconductor (NIS) tunnel junction pairs in series (SINIS junctions), both as a thermometer \cite{giazotto,panu} and as a heater \cite{leivo,meschkenature}. In contrast to our earlier measurements \cite{Zen,Maasilta2016}, where Cu was used as the normal metal, here we use Au-Ti bilayers as the normal metal. This is because Au-Ti was found to be a more robust material, in particular, it survived much better than Cu the contact to polymer resists and solvents in following lithography steps, increasing the fabrication yield.
More details on the fabrication and the measurement setup are given in the appendix.

\begin{figure*}[!htp]

\includegraphics[width=2\columnwidth]{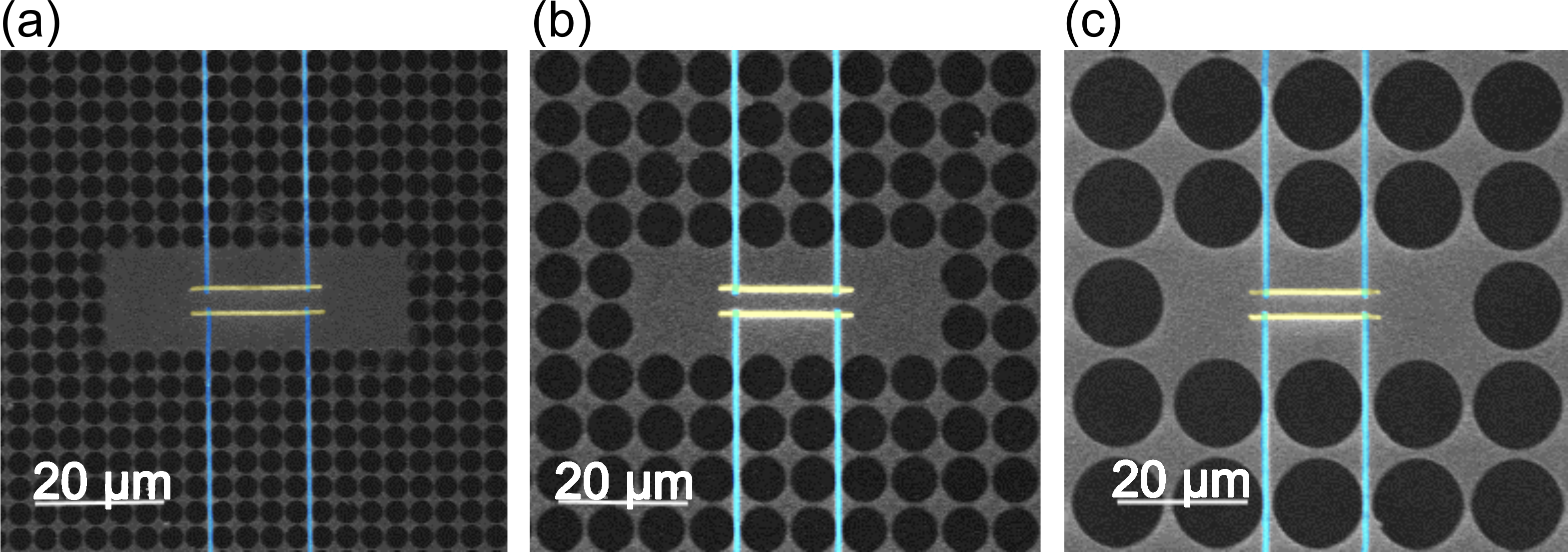}
\caption{\label{samples} SEM micrographs of the three different period SiN PnC structures measured, all with membrane thickness 300 nm and hole filling factor 0.7.   (a) $a= 4 \mu$m,  (b) $a= 8 \mu$m, and (c) $a= 16 \mu$m. Yellow shows the normal metal regions, blue the superconducting leads. } 
\label{threestructures}

\end{figure*}

The power vs. temperature curves were measured for the three different PnC structures with 4 $\mu$m, 8 $\mu$m and 16 $\mu$m periods, shown in Fig. \ref{threestructures}. As can be seen, the heater and thermometer geometry was identical in all samples, and the platform area without the holes about the same. Thus we have minimized the influence of other factors except the periodicity of the PnC.

\begin{figure*}[!htp]
\includegraphics[width=1.7\columnwidth]{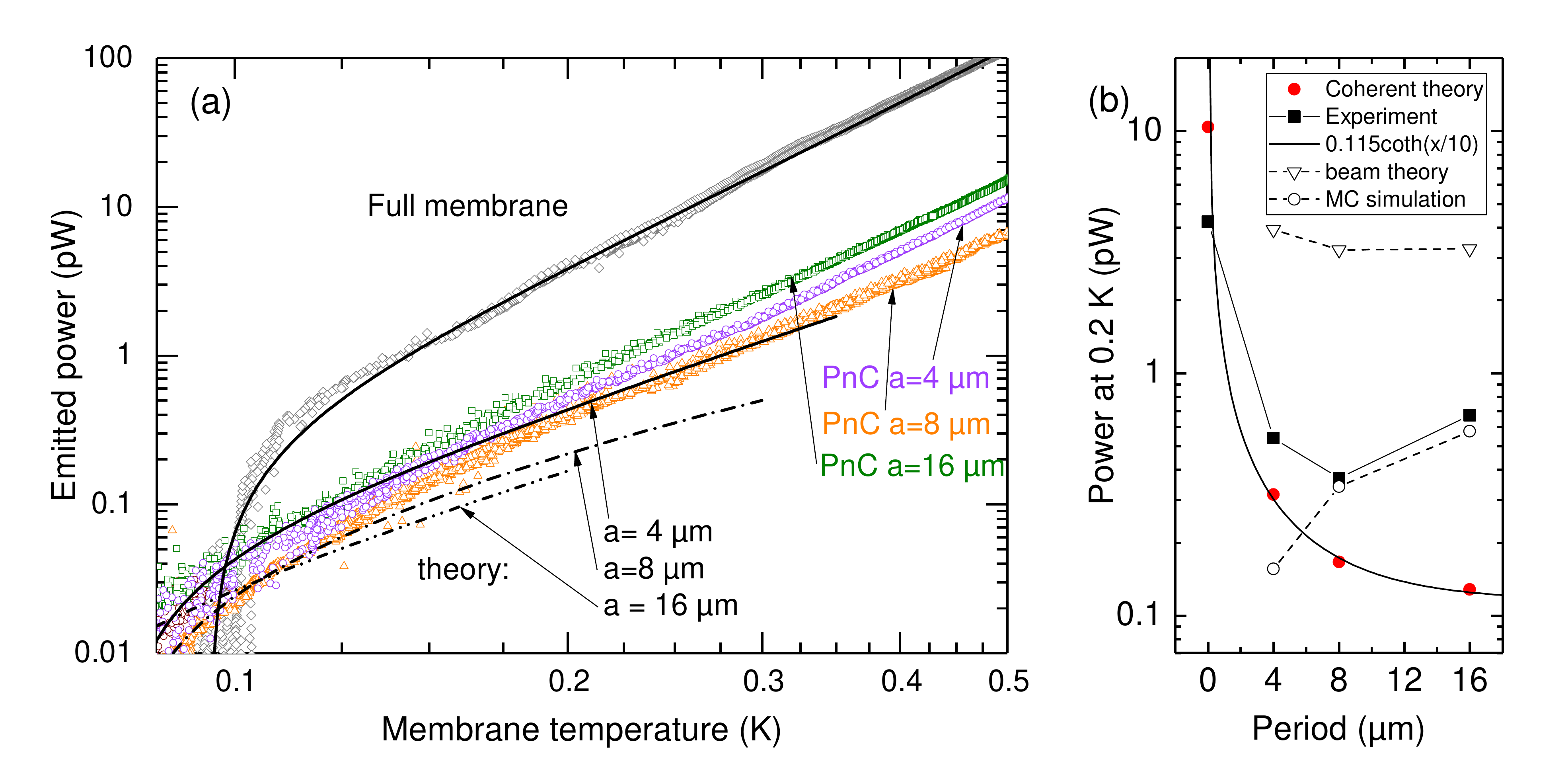}
\caption{(a) Measured power vs. temperature for the 4 $\mu$m (purple circles), 8 $\mu$m (orange triangles) and 16 $\mu$m (green squares) period PnCs and for an uncut membrane with same heater geometry (gray diamonds). Lines (solid, full membrane and 4 $\mu$m; dot-dash, 8 $\mu$; dot-dot-dash, 16 $\mu$) are theoretical curves calculated for each structure based on the ballistic model, Eq. \ref{PT}. (b) The observed power (black squares), the expected theoretical coherent power (red circles),  theoretical ballistic beam power (triangles), and MC simulation with partially diffusive scattering (open circles)  at a constant temperature of 0.2 K as a function of the PnC period. Zero period corresponds to full membrane. The line is a guide to the eye.}
\label{data1}
\end{figure*}

For the three different period PnC structures, the measured power vs. phonon temperature data is shown in Figure \ref{data1} (a). The membrane thickness (300 nm) and the hole filling factor (0.7) were kept constant. In addition, a measurement of an uncut membrane of the same thickness and with the same heater and thermometer geometry is shown.  The bath temperature was stable in each measurement, but varied between measurements in the range 55 mK - 78 mK for the PnC data, and was 95 mK in the membrane case.

Clearly, the emitted phonon power in all the PnC samples is about an order of magnitude below the uncut membrane, in other words thermal conductance is strongly reduced in the PnC structure. Interestingly, 
the 8 $\mu$m PnC has the lowest thermal conductance of all the measured samples, showing that there is a limit for how much it can be reduced simply by increasing the period with coherent band structure modification. For these type of devices, the optimum period for minimizing $G$ lies therefore between 4 $\mu$m and 16 $\mu$m. This ordering is in stark contrast with the calculations based on Eq (1), also shown in Fig. \ref{data1}, which predict that $G$ should keep decreasing with increasing period. Classically \cite{rayleigh} if the medium is diffusive, all the PnC structures would have the same thermal conductance, as it then only depends on the hole filling factor. On the other hand, in the boundary scattering limited models, an increase of $G$ with period is expected as neck size is increasing\cite{Lim,Verdier}. A non-monotonous behavior is thus an indication of breakdown of coherence and increased importance of boundary scattering.

For quantitative results, we define the thermal conductance as $G = P/(T_p-T_{\textrm{bath}})$, and get experimental values 
 4.5, 3.2 and 4.6 pW/K for the 4, 8, and 16  $\mu$m PnC structures 
at $T_p$ = 0.2 K. The PnC values are very low (comparable to a four-beam quantized thermal conductance \cite{PhysRevLett.81.232} of 3 pW/K at 0.2 K), with the 8 $\mu$m result the lowest observed specific conductance (independent of heater length) $G/L = 0.2$ pW/(K$\mu$m) for any PnC, so far (Table I). The values are also quite close to what was measured before \cite{Maasilta2016} for a 4 $\mu$m structure with a different heater/thermometer length and material. However, the membrane result is about an order of magnitude below what was observed in Refs. \cite{Zen,Maasilta2016}, so that the apparent reduction of $G$ now looks smaller. 
  This observation is not fully understood, but clearly results from the variations of the measurements in uncut membranes, and not of the PnCs. It is possible that $G$ in the membranes is much more prone to non-idealities, roughness variations, surface oxides, resist residues etc., and that the heater/thermometer geometry and material choice has a stronger effect in the uncut membranes. More work on this issue is clearly desirable.    

\begin{table}
\caption{\label{T2}Parameters of all square lattice PnCs measured to date. All samples have hole filling factor 0.7.  Here material refers to the heater and thermometer normal metal material, $a$ is the PnC period, $t$ the membrane thickness, $L$ the heater/thermometer normal metal length, and $G$ the measured thermal conductance.  }
\begin{tabular}{@{}ccccccc}
\hline 
Material  &    $a$   & $t$   & $L$ & $G(0.2 \textrm{K})$  & $G(0.2 \textrm{K})/L$ & Ref.    \\ 
			    & $\mu$m       & nm    &  $\mu$m       & pW/K & pW/(K$\mu$m) &  \\
\hline 
Cu     & 0.97          & 485    & 10      & 31.4  & 3.1    & \cite{Zen}       \\ 

Cu      & 2.4         & 485   & 10      &  11.6   & 1.2  & \cite{Zen}     \\ 

Cu      & 4         & 300   &   8    & 4.0   &  0.5 & \cite{Maasilta2016}    \\ 
 
AuTi      & 4        & 300    & 8      & 2.8    &  0.35 &       \\ 

AuTi     & 4         & 300    & 16      & 4.5   &   0.28 &     \\ 

AuTi     & 8         & 300   & 16      & 3.2    &  0.20 &     \\

AuTi     & 16        & 300    & 16      & 4.6    &   0.28 &      \\

\hline 
\end{tabular}   
\end{table}

In light of this, we have plotted the theoretical curves for the PnC structures (based on the FEM calculated band structures and ballistic power, Eq. \ref{PT})  in Fig. \ref{data1} (a) in such a way that the 4 $\mu$m data is fitted, and the 8 and 16  $\mu$m curves use the same absolute scaling factor. This way we can see that below 0.15 K, the 8 $\mu$m structure initially follows the thermal conductance reduction predicted by the theory, but later starts to deviate upward, as does the 4 $\mu$m data at a bit higher temperature around 0.2 K. On the other hand, the 16  $\mu$m structure does not follow the predicted reduction at all, and has about five times higher conductance around 0.2 K  than what the theory predicts.

Studying Fig. \ref{data1} (a) in more detail, we see that the temperature dependence of the emitted power 
is quite different between the uncut membrane and all the PnC samples, as was observed before \cite{Zen,Maasilta2016}.  For the membrane we have $P \sim T^{3.8}$ at the higher $T$ range, which is fully consistent with the ballistic Rayleigh-Lamb theory,  as can be seen by the nearly perfect fit of the theoretical curve. This is expected, as the cross-over from Rayleigh-Lamb modes to 3D bulk modes is estimated to take place at membrane thickness $\sim 1 \mu$m for SiN \cite{Kuhn2007}. In contrast, the temperature exponents are lower for the PnC structures, around 3.0 - 3.4. Comparing with the theory, we see that the temperature dependencies agree with the theory at the low temperature end, but start to deviate at higher temperatures.
Both the deviations of the magnitude and the temperature dependence thus lead to the conclusion that coherence can be destroyed by increasing the period beyond $\sim 8 \mu$m, or by increasing the temperature.

To make the deviations from the fully coherent theory more clear, we have also plotted the observed power and expected theoretical power at a constant temperature in Fig. \ref{data1} (b), as a function of the PnC period. The theoretical points are now presented without any fitting procedure, and take into account a geometrical correction due to the heater/thermometer dimensions (supporting information). We observe that for all the PnC devices the experimental power is higher than the coherent theory, which cannot be explained by adding a finite emissivity factor to Eq (1), as it could only reduce the amount of phonon emission from the heater. The experimental reduction between 4 and 8 $\mu$m PnCs follows the coherent theory trend, but the 16 $\mu$m data shows an increase rather than a decrease, which is again a strong indication of at least a  partial breakdown of the coherent picture. 

It is also possible to consider models (supporting information) where the coherence has been destroyed fully or partially. In the first model, we considered the case where the coherence length is longer than the cross-sectional neck dimensions, but shorter than the periodicity. In such a case the phonon modes carrying heat are defined by the set of quasi-1D beams leading out of the heated platform.
 Fig. \ref{data1} (b) also shows the expected power in that case, assuming no backscattering. We observe that the expected power level in such a model is an order of magnitude higher than the experimental observations, and thus describes the experimental data poorly.

Another model considers the full destruction of coherence, in which case phonons behave as particles, without the formation of new modes due to wave interference. Such a situation can be simulated with well known Monte Carlo (MC) techniques \cite{hori,LB}. We have performed such simulations (supporting information) with a code developed in Ref. \cite{lee}. As an example,  theoretical points based on a MC simulation with a specular scattering probability $p=0.875$ from the hole side walls are also shown in Fig. \ref{data1} (b). We observe that, indeed, such a model predicts the increasing trend seen for the longer period structures, in contrast to the prediction of the fully coherent theory. Although our model is crude and contains several unknown parameters, it gives support to the idea that the total conductance is a sum  of a coherent and an incoherent contribution. As the coherent contribution gets smaller with the periodicity and the incoherent one behaves in the opposite manner, a non-monotonous behavior can result, where at smaller periods the coherent theory dominates, but at the larger periods the incoherent theory is the more dominant contribution.                     

The evidence for partially diffusive boundary scattering can also be strengthened by pointing out that the observed side wall roughness scale ($\sim 7$ nm from SEM images such as Fig. \ref{samplefig1} (c)) indeed gives a finite diffusive scattering probability for the dominant thermal Lamb-wave modes in the temperature range of the experiment (Supporting information). Note that the observed rms sidewall roughness stays constant between the different samples, as the fabrication recipe stay the same. Thus, the observed effects are interpreted to arise from the direct effect of the geometry (neck size).

Finally, as the coherent theory is formulated in the ballistic regime, we can also build simple models (Supporting information) how the heater/thermometer geometry influences the measurement, in analogy with radiative photonic heat transfer problems. By modeling the heater/thermometer geometry with two parallel wires it is possible to derive analytical expressions for the view factors, set the power balance equations and solve for both the temperature of the heater and the thermometer as a function of dissipated power in the heater numerically.  This means that Eq (1), where only the heater temperature appears, can be extended to situations where heater and thermometer temperatures are not equal, with both heater and thermometer explicitly taken into account in the effective phonon emission into the PnC. For our geometry, the result is a correction term of magnitude $\sim 20$ \%, as discussed in supporting information.         

\begin{figure}[h]
\includegraphics[width=\columnwidth]{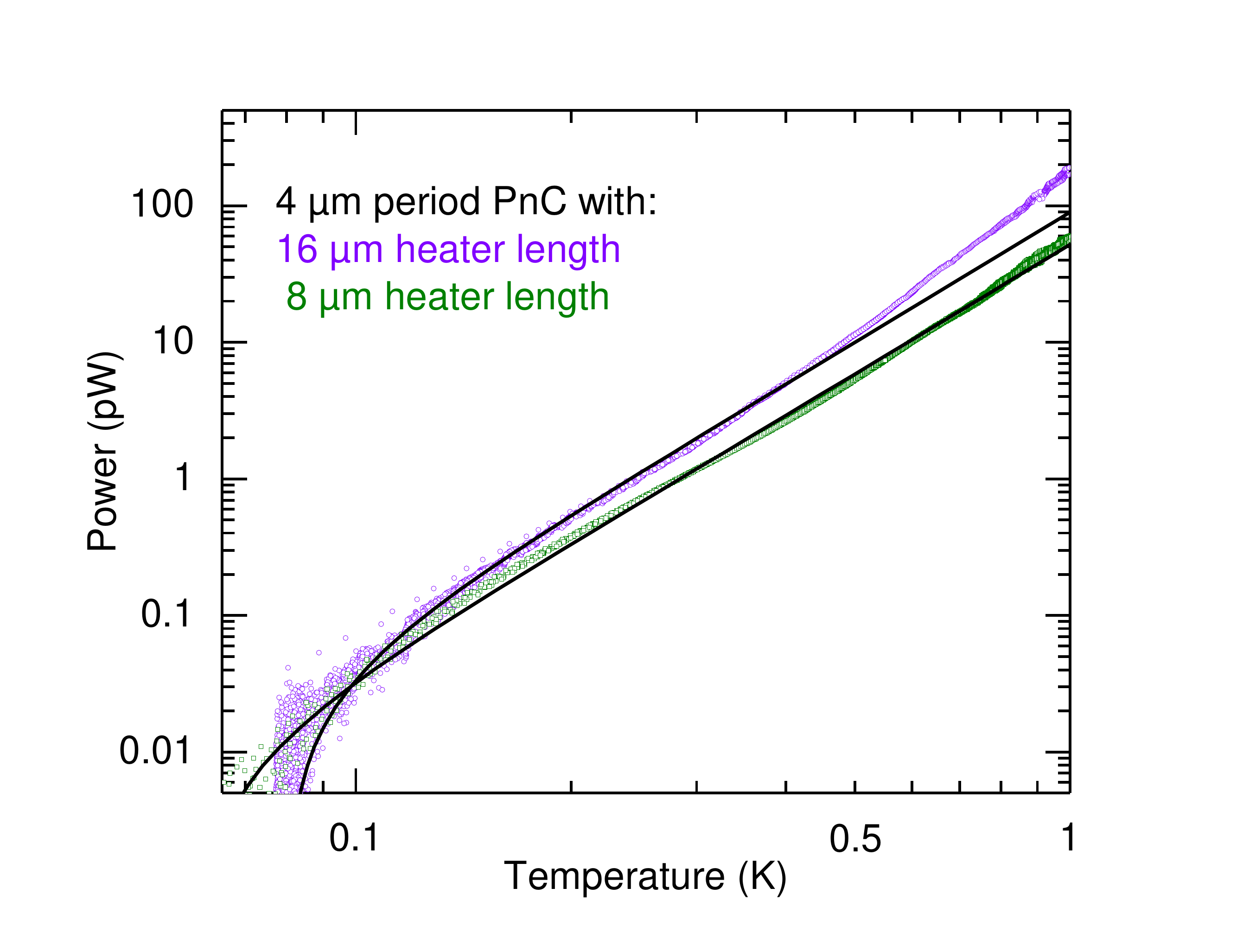}
\caption{Measured power vs. temperature for two 4 $\mu$m period PnCs, one with AuTi heater/thermometer length  8 $\mu$m (green squares) and the other 16 $\mu$m (purple circles). The theoretical line for the longer 16 $\mu$m heater structure is calculated by the parallel wire model, and used to determine the parameters (fit), whereas the line describing the 8 $\mu$m structure is calculated without fitting, by keeping all other parameters fixed and by just changing the wire length.} 
\label{data2}
\end{figure}

To test such modeling further, in Fig. \ref{data2}, we compare two measurements, both on a 4 $\mu$m period PnC, but with different lengths of the AuTi heater/thermometer structures, one with 8 $\mu$m, the other with 16 $\mu$m. In the diffusive limit, the heater and thermometer temperature should scale with $P/L$, the power/unit length of the heater, predicting a constant scale factor difference of two between the two measurements. Looking at the data, no such simple factor is observed, in contrast, the two measurements give quite similar results at low temperatures $T < 0.2 $ K, but deviate more strongly at higher temperatures. In contrast, the ballistic two-wire model explains the reduction of power in the shorter heater/thermometer structure very well, as shown by Fig. \ref{data2}. The partially incoherent ballistic 1D beam theory, on the other hand, predicts no dependence at all on the heater length, which does not agree with the experiment.

\section{Conclusions}

In conclusion, we have measured the thermal conductance of three two-dimensional square lattice hole array phononic crystal structures at sub-Kelvin temperature range, between periodicities 4 - 16 $\mu$m, which are much larger than in the previous work \cite{Zen, Maasilta2016}. The thermal conductance is still strongly reduced by the PnC structures, similar to Refs. \cite{Zen, Maasilta2016}, and we have demonstrated a record low specific conductance of  0.2 pW/(K$\mu$m) for 2D phononic crystals. However, the largest period does not give anymore the lowest thermal conductance as fully coherent theory predicts. The observed non-monotonous behavior as a function of the period can be explained by the interplay of coherent and incoherent phonon populations, with incoherent component becoming more dominant as the period grows. A practical limit thus seems to exist on what periodicity scale the coherent conductance dominates, depending on the details of the samples (such as surface roughness). 
 In our structures this limit 
was roughly at the 10  $\mu$m scale. 

In addition, temperature increase seems to start to destroy coherence as well. This is expected, since higher temperatures excite higher frequency phonons, which have stronger diffusive scattering probabilities than lower frequency phonons. In the future, if the two-dimensional specific thermal conductance $G/L$ needs to be reduced below 0.2 pW/(K$\mu$m) levels at 0.2 K, different geometries and strategies need to be employed, and the transition to the incoherent regime could be addressed further by studying the effect of the hole edge roughness and hole position disorder. The interpretation outlined in this work suggests that by reducing roughness, the range of validity of the coherent theory could be extended to larger periodicities and thus to smaller conductances. 

Our findings for the level of control and low values of thermal conductance are promising for applications in low-temperature ultrasensitive bolometric radiation detectors \cite{Irwin}, where coherent thermal control has the benefits of achieving the required low thermal conductance values with simple, compact and robust designs. This is in contrast to more traditional methods relying on diffusive surface scattering, where very long and fragile devices have to be used to obtain ultralow conductance values, making parameter control hard and high packing density a challenge.    

\begin{acknowledgments}


This study was supported by the Academy of Finland project number 298667 and the China Scholarship Council. We thank C. Dames and G. Wehmeyer for sharing the Monte Carlo code. Computational facilities provided by CSC- IT Center for Science Ltd are acknowledged.

\end{acknowledgments}

\appendix*

\section{Materials and methods}

\subsection{Sample fabrication}
All SINIS junctions were deposited on 300 $\mu$m thick single crystal (100) silicon wafers, coated on both sides with 300 nm thick low-stress silicon nitride films (Berkeley Nanofab). Silicon nitride membranes of about 300 $\mu$m by 300 $\mu$m area were fabricated by the standard KOH etching process, the same way as in Ref. \cite{Zen}, but excluding any tuning of the membrane thickness, i.e. the SiN membrane surfaces were kept in their native state. The heater and the thermometer SINIS junction pairs were then fabricated at the center of the SiNx membrane [Figure 1 (a)]. The distance between the heater and the thermometer was set to 4 $\mu$m for all samples, shorter than in Ref. \cite{Zen} but the same as in Ref. \cite{Maasilta2016}. The distance between the two superconducting leads of each SINIS pair was set to 16 $\mu$m, i.e. the heater length was about 16 $\mu$m, while it was half the length, 8 $\mu$m in Ref. \cite{Maasilta2016}. This heater length was adjusted to fit all the periods chosen, so that we could make a perfect PnC structure, avoiding any "phononic waveguide" features around the superconducting leads present in the older sample designs  \cite{Zen}. We do not have any indications that the possible waveguiding had any major impact in the older experiments, but when lower levels of thermal conduction are desired, it could start to influence the measurements.

The fabrication  of the SINIS junctions was similar to the process in Ref. \cite{Maasilta2016}, using electron-beam lithography. Al was used as the superconductor and deposited by a two-sided angle evaporation technique from a set of angles 70, 65 and 60 degrees for each side. This decreasing set of deposition angles was chosen to make sure each layer wholly deposit on top of the previous layer. The total Al thickness was 26.5 nm. Then, an in-situ AlOx tunnel barrier was formed by thermal oxidation at 200 mbar of oxygen for 5 mins.  

After a rotation of 90 degrees, a thin Ti adhesion layer was deposited from an angle of 70 degrees from both sides, with a total thickness 10 nm. Without breaking the vacuum, an Au layer of total thickness 37.3 nm was deposited the same way as Al. The purpose of this fairly complex high-angle deposition was to avoid the deposition of any normal metal layer on top of the superconducting leads, as they can influence phonon scattering. Finally after liftoff, two Al/AlOx/Ti-Au-Ti/AlOx/Al SINIS tunnel junction pairs with a junction area $\sim 370 \times 500$ nm$^2$ and a typical total tunneling resistance $\sim 10$ k$\Omega$ were obtained. Note that the tunneling resistances were typically lower than if Cu were used, even with similar oxidation parameters \cite{ilmo}.  

The PnC structures were fabricated after the SINIS junctions, with electron-beam lithography and reactive ion etching \cite{Zen}. A square lattice of circular holes was patterned on PMMA with a thin 7 nm layer of Al on top to prevent charging. After the removal of Al in NaOH and the development of the resist, it was used as an etch mask for reactive ion etching of the SiNx membrane, using a mixture of O$_2$ and CHF$_3$ in the plasma (ratio 10 to 1) at a power 100 W and a pressure 55 mTorr. The total etching time was divided into four steps (6 mins + 6 mins + 6 mins + 3mins 20 s) so that there was a cooling time in between each step. The SINIS junctions were located fully on the central uncut membrane island [Fig. 1(a)], and were therefore protected by PMMA during the plasma etching. From Figs. \ref{samplefig1} (b) and (c) it is clear that some non-idealities in the etching still remain. In particular, the side walls are not straight but sloped, and in the sample of Fig. \ref{samplefig1}, the neck dimension at the upper surface was $\sim 840$ nm, but at the bottom $\sim 1370$ nm. The design value for hole filling factor 0.7 in this case was 900 nm. Sloped side walls with similar roughness were also observed in the 8 $\mu$m and 4 $\mu$m period PnCs.  

\subsection{Thermal conductance measurement}
The measurement circuit is also shown in figure 1 (a). The heater was biased with a battery powered DC voltage source, which could be swept slowly. Both the heater voltage and current were recorded simultaneously during the bias sweep in a four probe configuration, using Ithaco 1201 and 1211 preamplifiers, respectively. Thus,  the emitted phonon power could be calculated as $P=IV$. Meanwhile, the thermometer SINIS pair was biased by a battery powered constant current source, and its voltage was measured with another Ithaco 1201 preamplifier as the dissipated power was swept. This voltage is proportional to phonon temperature $T_p$, and a calibration between voltage and temperature was performed by slowly sweeping the cryostat temperature without any heating power in the heater \cite{panu,Zen}. 

All measurements were performed in a $^3$He--$^4$He dilution refrigerator with a base temperature of $\sim$ 50 mK with several  
stages of filtering in the wires, because of the extreme sensitivity of the tunnel junction devices to unwanted spurious power loads. 
The setup used has pi-filters at the 4 K flange and a copper shielded Eccosorb filter box at the 1 K flange. From 1 K to 50 mK, a set of shielded superconducting wires are used. In addition,  twelve stand-alone copper shielded integrated Eccosorb+LC pi-filters with a measured  $> 80$ dB attenuation above at 40 MHz-10 GHz are used for each measurement line at the 50 mK sample stage thermalized to $T_{\textrm{bath}}$.



%

\newpage

\onecolumngrid

{\bf Supplemental information}

\maketitle

\section{Modeling of the effect of heater/thermometer geometry}

Here we describe the modeling for the effect of the parallel wire geometry (Fig. 1) on ballistic phonon transport, with the assumption of thermal equilibrium black body emission spectrum, which is isotropic.  First, one must find the view factor corresponding to the geometry in question. The 2D case of parallel thin wires  corresponds to the 3D geometry of infinitely long opposed parallel plates, and  the view factor $F_{12} = \sqrt{1+(d/L)^2}-d/L$ can be found in literature \cite{howell}, where $d$ is the distance between the wires and $L$ the length (Fig. 1). Thus, if we denote by $2I_1$ the radiative intensity (power per unit length into half-space) emitted by wire 1 {\em into an uncut membrane}, the power exchanged between the heater and thermometer is simply

\begin{figure}[h]
\includegraphics[width=0.1\columnwidth]{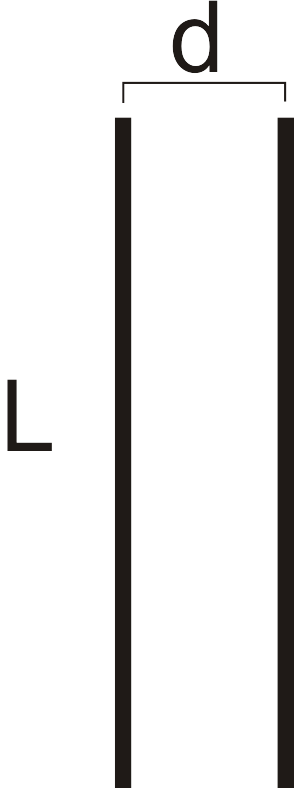}
\caption{The two wire geometry used in modeling.}
\label{samplefig1}
\end{figure}     

\begin{equation}
Q_{1\rightarrow 2} = 2I_1L\left(\sqrt{1+d^2/L^2}-d/L\right)  =  2I_1\left(\sqrt{L^2+d^2}-d\right),
\end{equation}    

where we also assumed unity black-body absorptivity of the thermometer wire, i.e, all incoming power is absorbed. Following the same definitions, we can write for the power emitted from the heater into the other half-space facing away from the thermometer $Q_{1\rightarrow {\rm half}} = 2I_1'L$ with $2I_1'$ the radiative intensity emitted by the heater wire 1 {\em into a phononic crystal} (PnC). In addition to these two contributions, there is one more emission contribution, which is the power emitted from a heater line segment towards the thermometer half-space, but escaping from the side openings in between the two wires directly into the phononic crystal, without having been intercepted by the thermometer wire. Because this leaked power and the power intercepted by the thermometer $Q_{1\rightarrow 2}$ must add up to the full hemispherical power, we get for the "leaked" power $Q_{1\rightarrow 3} = Q_{1\rightarrow {\rm half}} - Q_{1\rightarrow 2} = 2I_1L-2I_1(\sqrt{L^2+d^2}-d) = 2I_1 (L+d-\sqrt{L^2+d^2})$, where the last equation assumed membrane everywhere simply for the sake of the derivation of the "leaked" view factor $F_{13} = 1 +d/L - \sqrt{1+d^2/L^2}$. Thus, the power emitted this way into the phononic crystal, assuming minimal scattering from the PnC boundary, is 
\begin{equation}
Q_{1\rightarrow 3} = 2I_1'(L+d-\sqrt{L^2+d^2}). 
\end{equation}       

By symmetry arguments, similar terms exist for the emission from wire 2, the thermometer wire, where the intensities are of course different and denoted by $I_2$ and $I_2'$. Note also that this simple modeling does not take into account the anisotropic emissive properties of the PnC. Finally, there is incoming radiation from the bath (edge of the membrane), which is also approximated as a black surface. Due to the reciprocity of the view factors $A_iF_{ij}=A_jF_{ji}$, it is easy to see that the bath radiation has the same geometrical factors as the terms that were emitted into the bath, i.e. $Q_{bath \rightarrow 1,2} = 2I_{bath} (2L+d-\sqrt{L^2+d^2})$.  

With the help of the above derivations, we can now set the radiative balance equations for both wires, by equating incoming absorbed power with emitted power. Simplifying the notation a bit by defining temperature dependent "thermal conductance" (in reality power/unit emitter length) functions for the membrane, $f(T)$, and for the PnC, $g(T)$, we can write $I_1 = f(T_1)$, $I_2 = f(T_2)$, $I_1' = g(T_1)$, $I_2' = g(T_2)$ and $I_{bath} = g(T_{bath})$, and thus get 
\begin{eqnarray*}
P_{in} + 2f(T_2)(\sqrt{L^2+d^2}-d) + 2 g(T_{bath})(2L+d-\sqrt{L^2+d^2}) &=&  \\ 2g(T_1)L + 2g(T_1)(L+d-\sqrt{L^2+d^2}) + 2f(T_1)(\sqrt{L^2+d^2}-d) \\
2f(T_1)(\sqrt{L^2+d^2}-d) + 2 g(T_{bath})(2L+d-\sqrt{L^2+d^2}) &=& \\ 2g(T_2)L + 2g(T_2)(L+d-\sqrt{L^2+d^2}) + 2f(T_2)(\sqrt{L^2+d^2}-d),
\end{eqnarray*}
 where the first equation is for the heater with power input $P_{in}$, and the second for the thermometer. These equations simplify to the following two relations
\begin{eqnarray}
P_{in} &=&  (4L+2d-2\sqrt{L^2+d^2})[g(T_1) + g(T_2) - 2g(T_{bath})]   \\
f(T_1) &=& f(T_2) + [g(T_2) - g(T_{bath})]\left(\frac{2L+d-\sqrt{L^2+d^2}}{\sqrt{L^2+d^2}-d} \right),
\label{eqs}
\end{eqnarray}
which can be used to compute the power versus thermometer temperature curve $P_{in}(T_2,T_{bath})$, as the second equation can be used first to solve numerically for $T_1$ as a function of $T_2$ and $T_{bath}$. The conductance functions $f$ and $g$ are either computed numerically using Rayleigh-Lamb theory for full membranes and FEM simulations for the PnC structures \cite{Zen}, [Fig. 4(b), main text], or $f$ and $g$ can be treated as power-law functions fitted into the experimental data [Fig. 5, main text].   

For the case where the PnC conductance $g$ is much smaller than the membrane conductance $f$, we see that the heater and thermometer temperatures are approximately equal $T_1 \approx T_2$, and the power vs temperature relation simplifies to
\begin{equation}
P_{in} \approx 4L\left(2+d/L-\sqrt{1+(d/L)^2}\right) [g(T_1) - g(T_{bath})].
\label{approx}  
\end{equation} 
This expression has the correct single wire limit when $d/L \rightarrow 0$: $P_{in} \rightarrow 4L [g(T_1) - g(T_{bath})]$.  

For the case of the experiments here, the condition $g(T) << f(T)$ is quite well satisfied, and thus Eq. \ref{approx} is a good approximation. We see, however, that even in this limit, there is a geometrical correction to the single wire case, given by the factor in parentheses in Eq. \ref{approx}. This factor has a value 1.1875 in the geometry of the three PnC structures where $d/L = 1/4$ (1.382 for the 8 $\mu$m long wire with $d/L=1/2$), meaning a $\sim 20$ \% ($\sim 40$ \%) increase in the apparent conduction. However, full equations (\ref{eqs}) were used in the calculations presented in the main text.   

In Figure 2, we show an example of the numerically calculated emission power for the $ 4 \mu$m period PnC structure, compared with the bare theory where only single heater wire is assumed, using FEM calculated band structure and the ballistic theory presented above. It is clear that the simple theory with constant increase factor, Eq. \ref{approx}, works well above the bath temperature. 
 
\begin{figure}[h]
\includegraphics[width=0.7\columnwidth]{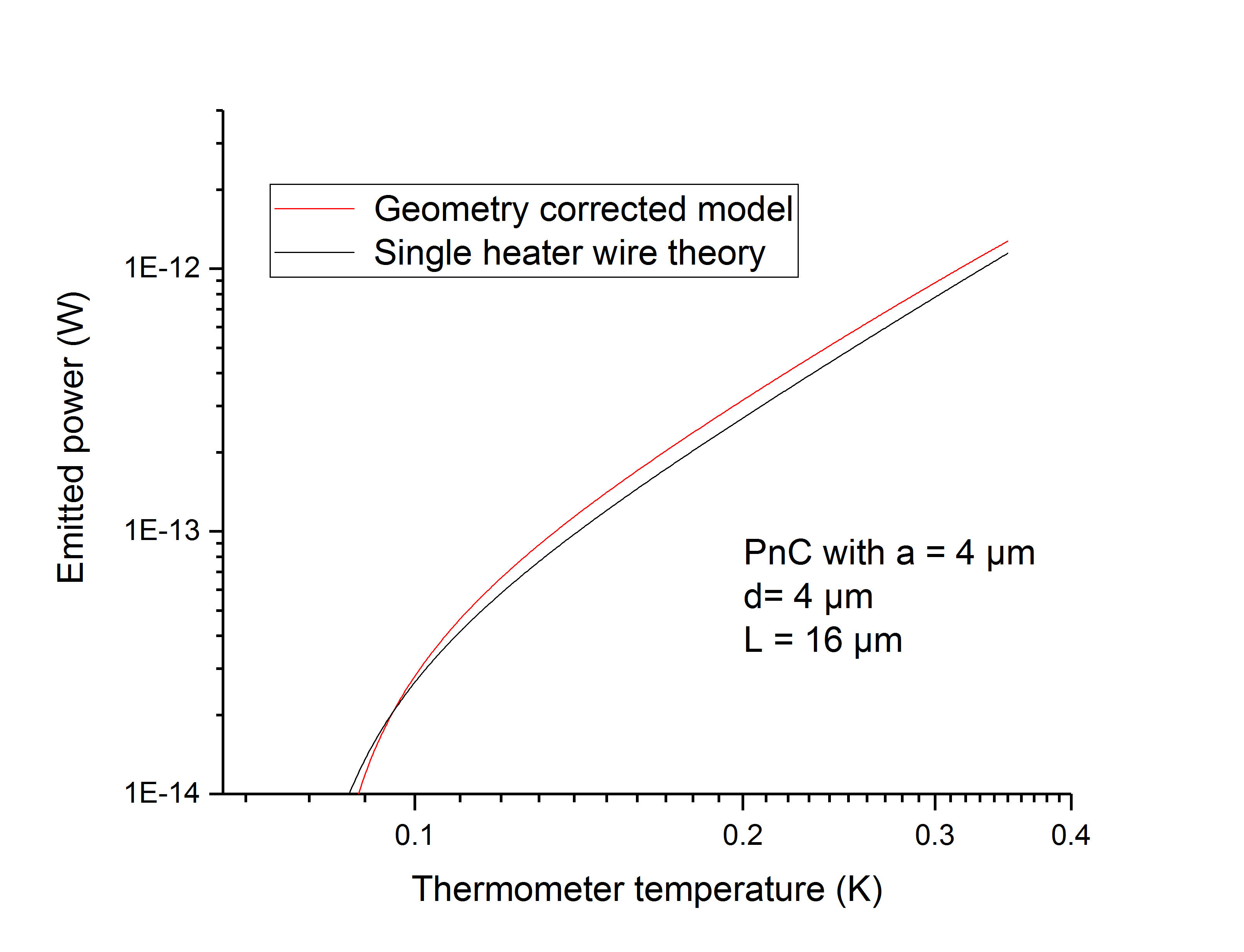}
\caption{Comparison of the geometry corrected computational two-wire result to the one with single wire for a 4 $\mu$m period PnC structure.}
\label{samplefig2}
\end{figure}     

\newpage

\section{Modeling of the heat flow as set of ballistic beams}

Assuming an average speed of sound 5000 m/s, roughly 80\% of the phonon modes at $T = 200$ mK have wavelengths shorter than 400 nm, which is 10-40 times smaller than a single period in the PnC samples. Thus, it is natural to question if coherence, particularly for the high frequency phonons, can extend over such length scales. 

If the coherence in the 2D phonon system in the PnC collapses but the sidewalls are smooth, the PnC could be thought of as a network of one dimensional thin ballistic beams - those are the interconnected bridges between the holes. In the simplest of those models, we assume that the phonons traveling outward from the heater cross one of the bridges in the first layer of PnC holes and never scatter back to the heater, thus only the first layer of beams defines the thermal conductance. For the shorter wavelength phonons, these bridges can then be modeled as infinitely long beams with constant cross-sectional shape. We thus neglect any frequency dependent scattering caused by changing geometry or interfaces, and assume a unity transmission for all modes.

The number and the width of first layer bridges varies over the three PnC samples with different periodicities, which has an effect on the phonon power. It is clear that at $T=200$ mK we are well above the quantum limit with these geometries, and the higher order phonon modes (5th, 6th, etc.) have to be taken into account (see Table \ref{tbl1D}). Therefore, 1D approximations, such as the ones used in \cite{PhysRevLett.81.232}, are not sufficient, and we have to numerically calculate the phonon spectrum for each beam geometry using the finite element method.

\begin{figure}
\includegraphics[width=0.5\columnwidth]{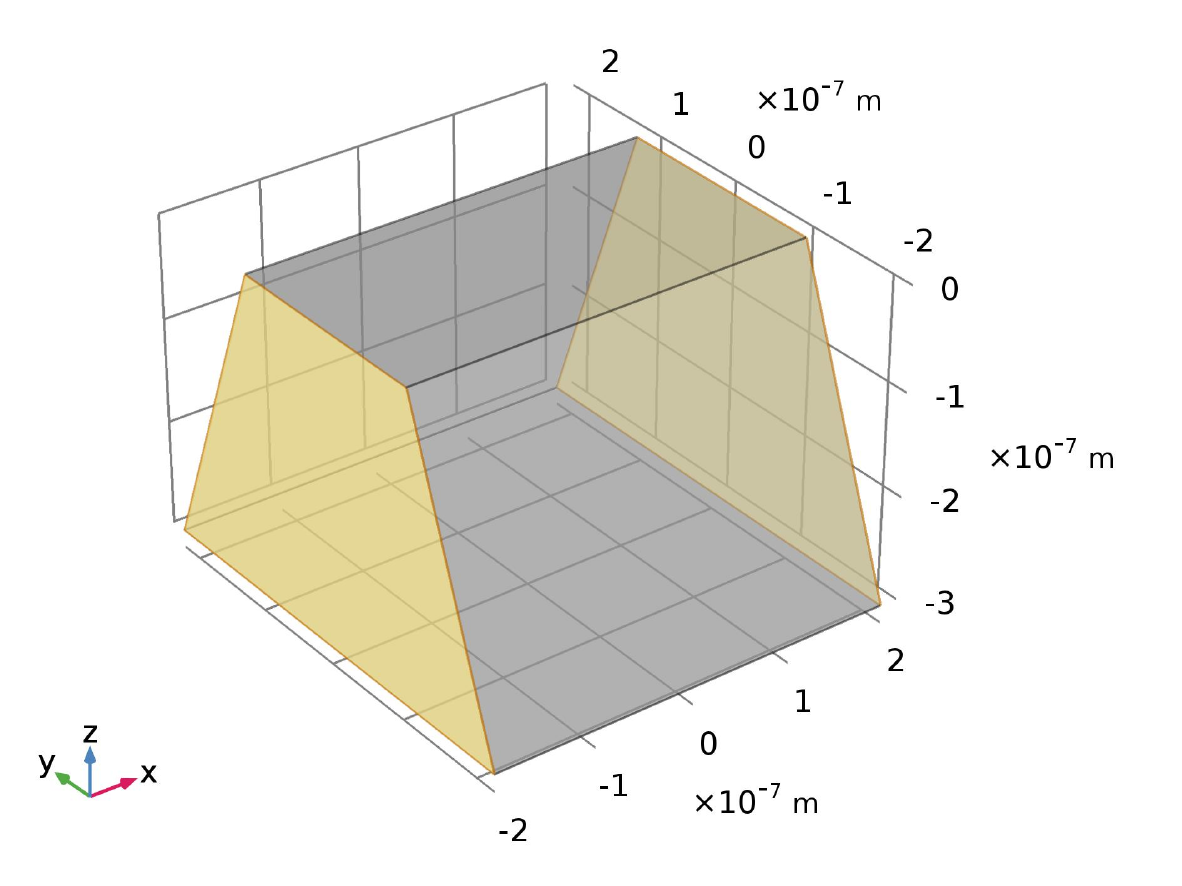}
\caption{Model of an infinitely long trapezoidal beam where the periodic Bloch boundary condition is applied on the highlighted boundaries for the calculation of phonon eigenfrequencies.}
\label{trapzbridge}
\end{figure}

\begin{figure}
\includegraphics[width=0.7\columnwidth]{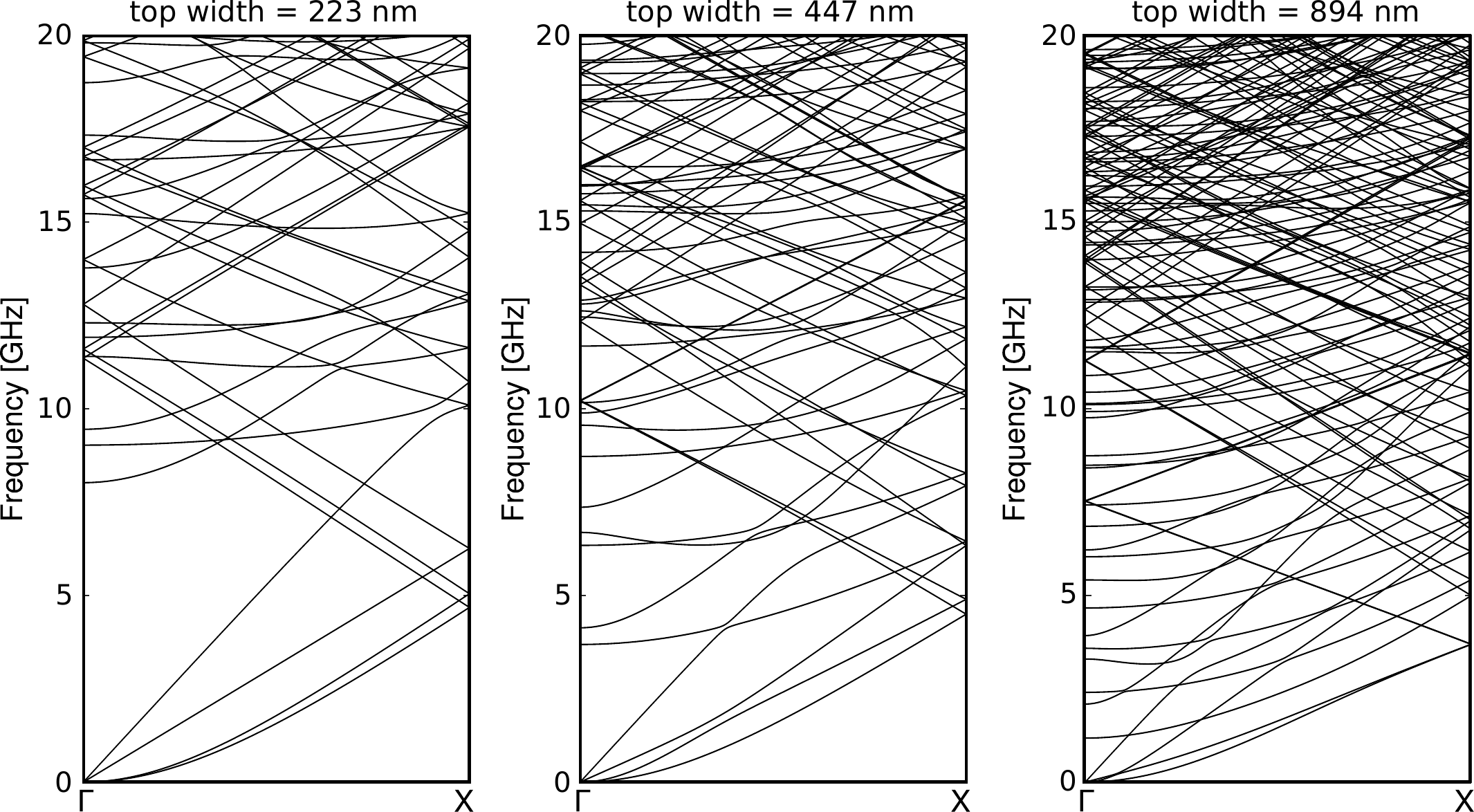}
\caption{Numerically calculated phonon eigenfrequencies for the trapeziodal quasi-1D wires corresponding to the three PnC periodicities.}
\label{spectra}
\end{figure}

The experimental cross-sectional shape of the bridge (see Fig. 1(c), main text) is approximated as an isosceles trapezoid (Fig. \ref{trapzbridge}) where the bottom width is set to  two times the top for all designs, and the top widths are calculated to match the filling factor design value of 0.7. Using such a geometry, we then calculate the phonon spectra for each PnC geometry, as shown in Fig. \ref{spectra}.

\begin{table}
\centering
\caption{Beam dimensions and cut-off frequencies of the higher order phonons.}
\label{tbl1D}
\begin{tabular}{l|r|r|r}
PnC lattice const [um] & 4 um & 8 um & 16 um \\
\hline
Number of bridges & 36 & 20 & 12 \\
Bridge top width [nm] & 223 & 447 & 894 \\
5th mode cutoff [GHz] & 8.0 & 3.7 & 1.2 \\
6th mode cutoff [GHz] & 9.0 & 4.1 & 2.1 \\
7th mode cutoff [GHz] & 9.5 & 6.3 & 2.4 \\
8th mode cutoff [GHz] & 11.4& 6.7 & 3.3 \\
\end{tabular}
\end{table}
 
Finally, for each wire, the quasi-one-dimensional ballistic phonon power  $P_{\text 1D}$ at temperature $T$ is evaluated by integrating over the energies $\hbar\omega_j$ and velocities $d\omega_j/dk$ of the out-going phonon modes $j$, taking into account the thermal population $n(\omega, T)$:
\begin{equation}
P_{\text 1D}(T) = \frac{1}{2\pi}\sum_{j}\int_{K^+}dk\,\hbar\omega_j\,\left|\frac{d \omega_j}{d k}\right|\, n(\omega_j, T).
\end{equation}

\begin{figure}
\includegraphics[width=0.7\columnwidth]{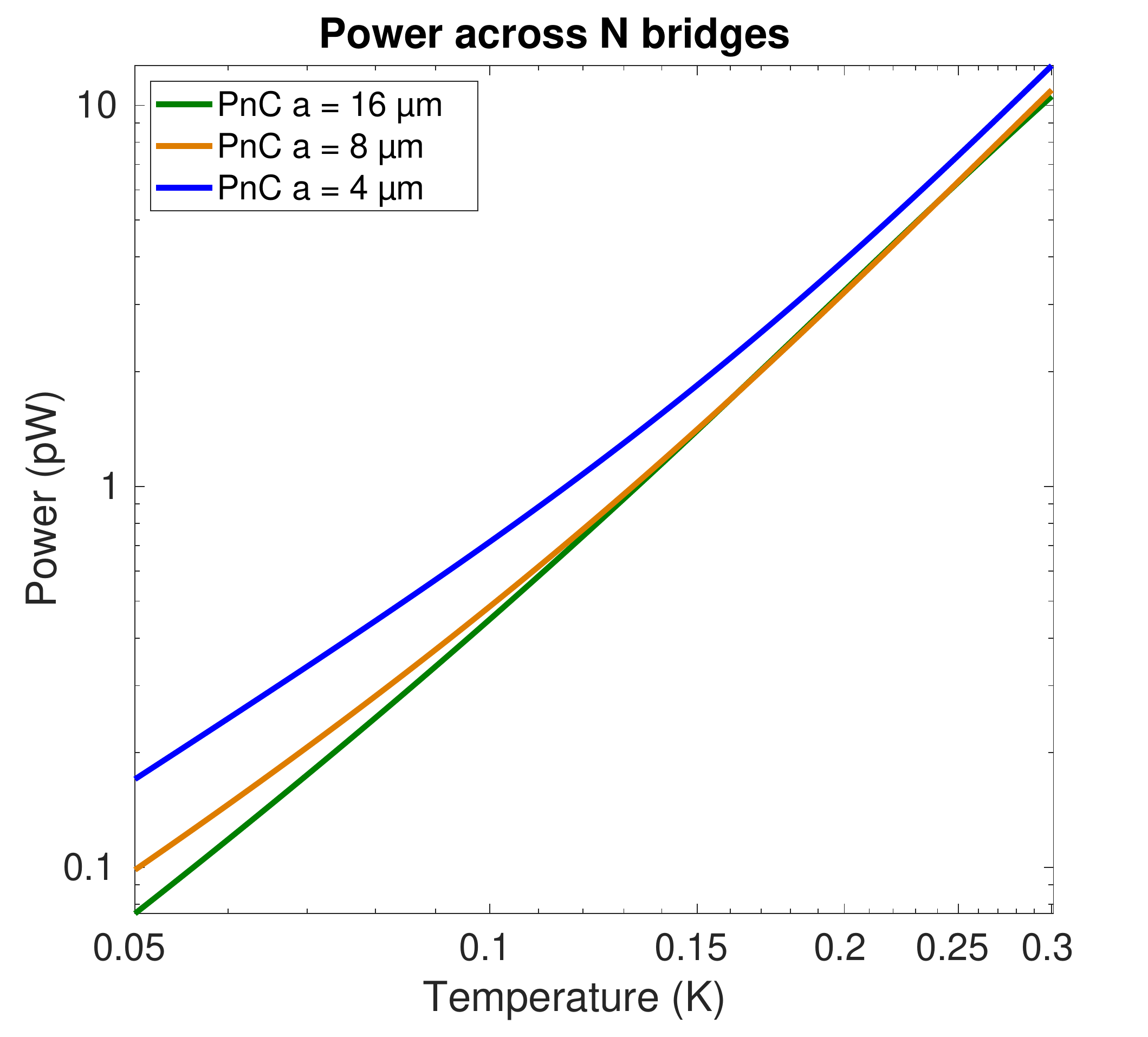}
\caption{Numerically calculated ballistic power carried by the trapeziodal quasi-1D wires corresponding to the three PnC periodicities.}
\label{power}
\end{figure}

By taking into account the varying number of beams supporting the heater/thermometer platform in each PnC sample (Table \ref{tbl1D}), we can finally compute the total emitted power of each set of beams, getting a result shown in Fig. \ref{power}. Although a single wider beam carries more power than a narrower one, the total power does not change much as the number of beams goes down with increasing PnC period, as well (Table I).   

\newpage

\section{Monte Carlo simulation of the incoherent boundary scattering limited conductance}

\begin{figure}
\includegraphics[width=0.7\columnwidth]{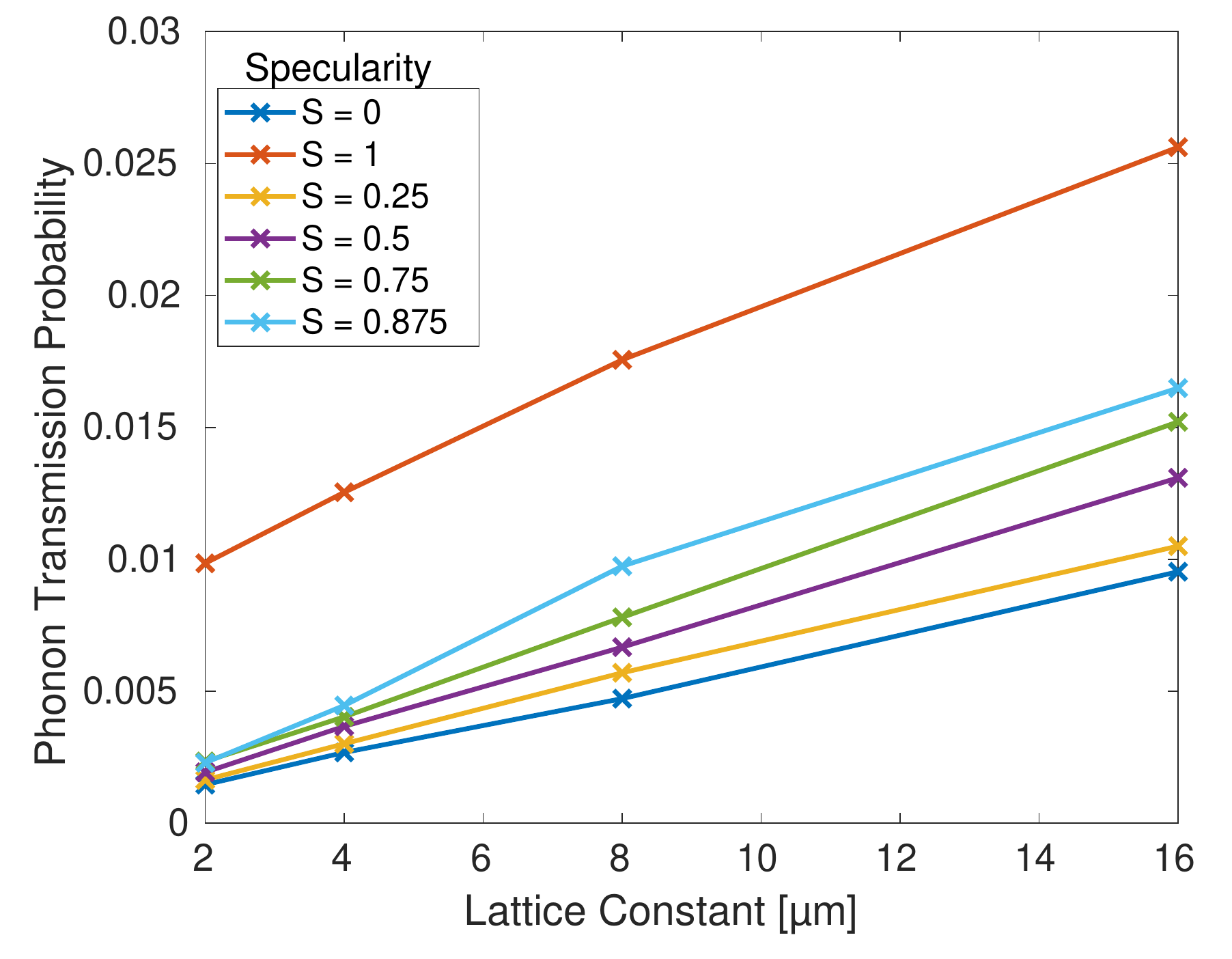}
\caption{Calculated average transmission probability of incoherent phonons through a square-lattice circular hole PnCs of varying period and probability of specular scattering $S$ from the hole side walls.  Hole filling factor 0.7, membrane thickness 300 nm, width of the PnC 64 $\mu$m.}
\label{MC}
\end{figure}

If coherence is fully destroyed, a model calculation for the thermal conductance through a phononic crystal structure can be performed based on the particle picture of phonons, using the Monte Carlo technique  and the Landauer-B{\"u}ttiker formalism \cite{hori,LB}. We have applied the code developed in Ref. \cite{lee} (where full details of the computational implementation can be found) to calculate the average phonon transmission probabilities $\left\langle \tau \right\rangle$ through square lattice circular-hole phononic crystals with varying periods $2 - 16 \mu$m, corresponding to the experiment in the main text (membrane thickness 300 nm, hole filling factor 0.7), see Fig. \ref{MC}. The length of the simulated structure was kept constant at 64 $\mu$m. No bulk scattering was included, and the top and bottom surfaces were assumed to be fully specular ($S = 1$), whereas for the side walls, specular scattering probabilities were varied between the fully diffusive ($S = 0$) and the fully specular ($S = 1$) limits.   

As we see, for all values of the specularity parameter, the transmission increases with period, leading to the increase of thermal conductance, as the conductance is directly proportional to $\left\langle \tau \right\rangle$ \cite{lee}. This can be understood by the geometrical argument of the increasing neck size.   

We note that the code used assumes 1D heat flow, and is therefore an approximation to the full geometry of our experiment, where heat spreads out two-dimensionally into the 2D PnC structure. However, as we are not interested in the exact numbers in the context of this paper, but only want to discuss the trend, the simpler simulation is adequate.    

\section{Estimation of the effect of boundary roughness at sub-Kelvin temperatures}

To give further support for the picture, whether boundary scattering can start to affect the coherent thermal conduction at sub-Kelvin temperatures in thin membranes, one can do a simple estimate based on the concept of the dominant thermal wavelength, if a model for the wavelength dependence of the specular scattering probability is in place. A simple model for the specularity was given by Ziman \cite{Ziman}, where the specular scattering probability $S$ depends on the wavelength $\lambda$ of the scattering phonon as
\begin{equation}
S(\lambda) = \exp\left[ -16\pi^3\left(\frac{R}{\lambda}\right)^2\right],
\label{Zimanmodel}
\end{equation}
 where $R$ is the rms roughness. 

\begin{figure}
\includegraphics[width=0.7\columnwidth]{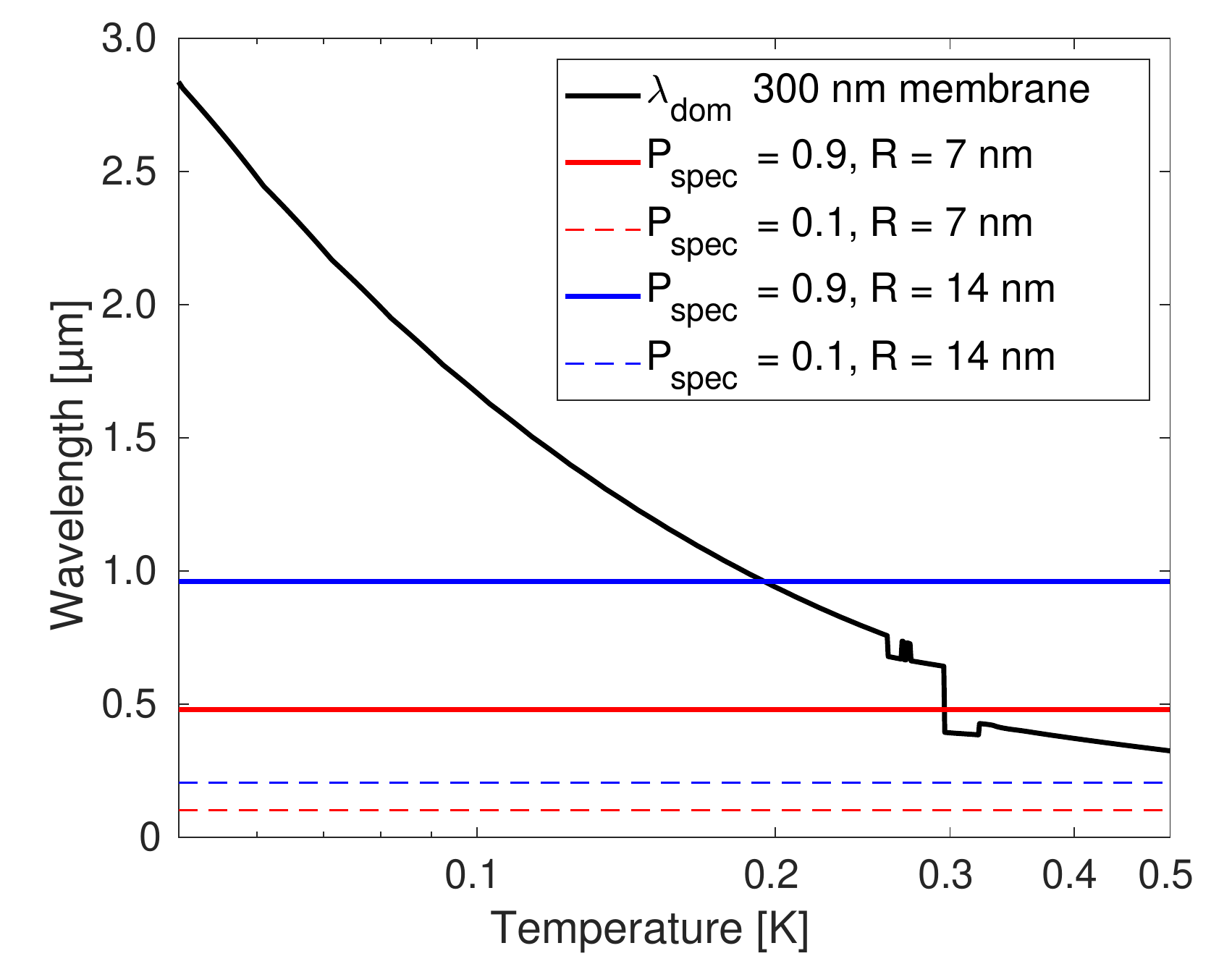}
\caption{The dominant wavelength of the Lamb-waves of a 300 nm thick membrane vs. temperature (solid black line), the wavelength producing $S=0.9$ (red and blue solid lines) or $S=0.1$ (red and blue dashed lines) in the Ziman model of Eq. \ref{Zimanmodel}. }
\label{lambdadom}
\end{figure}

In Fig. \ref{lambdadom} we compare the numerically calculated dominant wavelength of the phonon eigenmodes in thin membranes (Lamb modes \cite{graff}) to a few examples of wavelength values that start to give a sizable probability for diffusive scattering ($S=0.9$), for two realistic values of the roughness parameter $R=7,14$ nm.  
We see from the plot that, for example, the experimentally estimated value of roughness of 7 nm produces 10 \% diffusive scattering already at 300 mK, which is a temperature scale roughly in line with  the range, where the experimental temperature dependence starts to deviate from the fully coherent theory (Fig. 4(a) in the main text.) 

The dominant wavelength was defined with respect to the most power carried by membrane phonons $\omega_j$, $j = 1,2,\ldots$. We first calculated the spectral power $P(\omega)$, with
\begin{equation}
P(\omega) = \sum\limits_{j} 2\pi|k_j(\omega)|\cdot|v_j(\omega)|\cdot n(\omega,T)\cdot \hbar\omega
\end{equation}
where $k_j(\omega)$ is the inverse of the phonon dispersion branches $\omega_j(k)$ (where defined),
$v_j = \partial\omega_j/\partial k$ is the group velocity and $n(\omega,T)$ is the Bose-Einstein distribution. For each temperature $T$, we then search for the maximum $P(\omega)$ and convert the corresponding angular frequency $\omega$ to wavelength by finding the average wavelength of the membrane phonons as a function of $\omega$ 
\begin{equation}
\tilde{\lambda}(\omega) = \frac{1}{L}\sum\limits_{j} \frac{2\pi}{|k_j(\omega)|}2\pi k_j(\omega),
\end{equation}
 where $L = 2\pi\sum_j k_j(\omega)$.


%

\end{document}